\documentclass{rspublic}
\usepackage{graphicx}
\usepackage{bm}

\begin{document}

\title[Lagrangian views on turbulent mixing of passive scalars]
{Lagrangian views on turbulent mixing of passive scalars}

\author[K. R. Sreenivasan \& J. Schumacher]{Katepalli R. Sreenivasan$^1$ \& J\"org Schumacher$^2$}

\affiliation{$^1$The International Centre for Theoretical Physics \\Strada Costiera 11, 34014 Trieste \\
$^2$Institut f\"ur Thermo- und Fluiddynamik, Technische Universit\"at Ilmenau, Postfach 100565, D-98684 Ilmenau, Germany}

\label{firstpage}

\maketitle

\begin{abstract}

The Lagrangian view of passive scalar turbulence has recently
produced interesting results and interpretations. Innovations in
theory, experiments, simulations and data analysis of Lagrangian
turbulence are reviewed here in brief. Part of the review is closely
related to the so-called Kraichnan model for the advection of the
passive scalar in synthetic turbulence. Possible implications for a
better understanding of the passive scalar mixing in Navier-Stokes
turbulence are also discussed.
\end{abstract}


\section{Introduction}
Eulerian and Lagrangian points of view on the motion of fluids, and
on the mixing of scalars, are equivalent (Prandtl \& Tietjens 1934)
but cannot be related to each other in analytically tractable ways
except in a few special instances. The bridge between the two
descriptions is formally obvious, at least for a conserved scalar. A
batch of a massless tracer particle which is advected in an
incompressible turbulent flow ${\bf u}$, subject to the diffusion
$\kappa$, follows the Langevin equation (Gardiner 2004) given by
\begin{equation}
\dot{\bf x}={\bf u}({\bf x}(t),t)+\sqrt{2\kappa}\,{\bm\chi}(t)\,,
\label{sde1}
\end{equation}
where ${\bm \chi}(t)$ is vectorial white noise that is
statistically independent in each of its three components. This
equation is complementary to a Fokker-Planck equation for $\theta({\bf
x},t)$, the probability density function (PDF) of the tracer at time
$t$ at position ${\bf x}$. The equation
\begin{equation}
\frac{\partial\theta}{\partial t} + ({\bf u}\cdot{\bf \nabla})\theta
=\kappa{\bf\nabla}^2\theta\ \label{Fokker}
\end{equation}
is exactly the advection-diffusion equation for a scalar field in
the Eulerian frame. The scalar diffusivity is denoted by $\kappa$.
Its ratio to the kinematic viscosity $\nu$ of the fluid is the
Schmidt number
\begin{equation}
Sc=\frac{\nu}{\kappa}\,,
\end{equation}
or, for an advected temperature field, the Prandtl number, $Pr$.

The scalar mixing process at a given Reynolds number is
significantly different in the Kolmogorov-Obukhov-Corrsin regime
given by $Sc\le 1$ (Kolmogorov 1941, Obukhov 1949, Corrsin 1951)
from that in the Batchelor regime given by $Sc\gg 1$ (Batchelor
1959). While the scalar is advected by the turbulent inertial range
in the former, it is mainly advected in the latter by smooth
velocity structures of the viscous range of turbulence.

In turbulent mixing of passive scalars, and in turbulence in
general, Eulerian and Lagrangian views have their specialized
places. For example, the former appears better suited for studying
problems such as nonpremixed combustion in reacting flows (Lin\~{a}n
\& Williams 1993, Peters 2000). Indeed, the comparative ease of
making Eulerian measurements renders them more appealing in a number
of instances. If, on the other hand, one has to calculate the
dispersion of $\theta$ arising from a few localized sources, the
Lagrangian view is more appropriate (Sawford 2001). Examples of such
applications include fumigation (Sawford {\it et al.} 1998), the
spread of buoyant plumes (Heinz \& van Dop 1999), and the surface
motion of buoys over the sea (Maurizi {\it et al.} 2004). Lagrangian
studies are being recognized as increasingly important in
geophysics, e.g. for the mixing of plankton and other biomatter in
the upper ocean (Seuront \& Schmitt 2004) and for droplet dynamics
in cumulus clouds (Vaillancourt {\it et al.} 2002).

In spite of their wide-spread utility, Lagrangian data are harder to
obtain experimentally and are more expensive to compute. For
example, it is unclear if we can ever measure with satisfactory
accuracy high-order moments of the particle acceleration along a
trajectory in a turbulent flow of moderately high Reynolds number
(Yakhot \& Sreenivasan 2005); see also discussion of Fig.\ 1 in
Sec.\ 3. The relevant point here is that measuring (as an example)
the fourth moment of the Lagrangian acceleration is equivalent to
measuring the twelfth moment of Eulerian velocity differences (which
has not been obtained so far with impeccable accuracy). The basis
for this connection is the relation for the Lagrangian acceleration
expressed in terms of Eulerian velocity increments,
\begin{equation}
a\approx\frac{(\delta_{\eta}u)^2}{\eta}
\approx\frac{(\delta_{\eta}u)^3}{\nu}\,,
\end{equation}
where $\eta$ is the fluctuating small scale around the Kolmogorov
length $\eta_K$ and the combination $\eta \times \delta_{\eta}u/\nu
= 1$, relating the velocity increment $\delta_{\eta}u$ on the length
scale $\eta$ to $\eta$. The difficulties are compounded by the
tendency of most conventional Lagrangian tracers to cluster near
solid walls, which produces a large mismatch of information between
fluid particles and tracers. Large shear will have similar effects.
Thus, one should ask if the additional complexities associated with
Lagrangian calculations and measurements justify the investment and
effort, leaving aside their relative novelty for the present: When
is the Lagrangian perspective the method of choice? What makes it
advantageous? What have we learnt in the big picture?

New insights into these questions, especially on the scalar mixing
problem, have become possible for the so-called Kraichnan model
(Kraichnan 1968, 1994). For a review with an extensive list of
references, see Falkovich {\it et al.} (2001). The Kraichnan model
has been applied to diverse physical circumstances such as the
mixing of passive scalars and their dynamics in compressible
turbulence (Gaw\c{e}dzki \& Vergassola 2000, Bec {\it et al.} 2004).
In a further development, the claim has been made (with certain
careful caveats) that the anomalous scaling of active scalars can be
understood in terms of passive scalar fields (Ching {\it et al.}
2003).

All these developments deserve some comment. This is one of our
purposes here. Secondly, we will comment on these theoretical
achievements in the context of numerical and experimental work on
the mixing of the passive scalar in Navier-Stokes turbulence. In the
process, we draw attention to a few open questions that should be
addressed in the future. We pay particular attention to the
Kraichnan model and its relation to homogeneous turbulence. Clearly,
there are many other aspects related to the turbulent mixing which
cannot be covered here. For studies of turbulent mixing in
inhomogeneous shear layers, jets or the mixing by Rayleigh-Taylor
instabilities, see, for example, the review by Dimotakis (2005) or a
recent work by Cabot \& Cook (2006).

The outline of the paper is as follows. We will briefly review the
scalar advection in a white-in-time Kraichnan flow in the next
section and consider basic ideas which are closely tied with the
Lagrangian view. We then discuss numerical and experimental efforts
connected to the Lagrangian picture of mixing in Navier-Stokes
turbulence. In the last section, we relate these findings to passive
scalar mixing in Navier-Stokes turbulence, particularly for high
Schmidt numbers.

\section{Scalar mixing in white-in-time turbulence}
It is useful to begin with the anomalous scaling for the passive
scalar obeying the Kraichnan model, in large part because some exact
results are available. The crucial element of the model is that it
uses for the velocity in the advection-diffusion equation a
stochastic Gaussian field with a time correlation that decays
infinitely rapidly (or is ``white in time") and a spatial
correlation that has a power law structure with a prescribed scaling
exponent, $0< \zeta<2$. That is,
\begin{equation}
\langle u_i({\bf x},t) u_j({\bf y},t^{\prime})\rangle=D_{ij}({\bf
x}-{\bf y}) \delta(t-t^{\prime})\,,
\end{equation}
with
\begin{equation}
D_{ij}({\bf r})=D_0\delta_{ij}-d_{ij}({\bf r})=
D_0\delta_{ij}-D_1\left((2+\zeta)\delta_{ij}-\zeta\frac{r_i
r_j}{r^2}\right)r^{\zeta}\,,
\label{Kraichnandiffusivity}
\end{equation}
where ${\bf r}={\bf x}-{\bf y}$, $r=|{\bf r}|$ and $i, j=1, 2, 3$,
and $\langle . \rangle$ denotes a suitable average; $D_{ij}$ is a
diffusivity with the dimension of $L^2T^{-1}$, and $D_0$ and $D_1$
are constants. The case $\zeta=0$ stands for advection in a very
rough flow and $\zeta=2$ for transport in a smooth flow (see Bernard
(2000) for a compact introduction to the subject). This power-law
scaling is similar to the Navier-Stokes case (though the scaling
exponent $\zeta$ here assumes an arbitrary value between 0 and 2),
but the temporal scaling is qualitatively different. For statistical
stationarity, a random forcing $f_{\theta}({\bf x},t)$ has to be
added to the right hand side of (\ref{Fokker}), with the property
that
\begin{equation}
\langle f_{\theta}({\bf x},t) f_{\theta}({\bf
y},t^{\prime})\rangle=C(r/L) \delta(t-t^{\prime})\,.
\end{equation}
The function $C(r/L)$ varies only on the large scale $L$ and decays
rapidly to zero for smaller scales. Kraichnan's insight was that
this model possesses the essential elements of the scalar mixing
while retaining analytical tractability. Indeed, it has been
possible to establish anomalous scaling (Falkovich {\it et al.}
2001) for this model even though the idealized advecting flow itself
does not exhibit such anomaly. In other words, the scaling exponents
for the Eulerian structure functions $S_n(r)=\langle(\theta({\bf
x}+{\bf r}) - \theta({\bf x}))^n\rangle\sim r^{\xi_n}$ of the scalar
increments differ from the classical Kolmogorov-Obukhov form (which
is $\xi_n=(2-\zeta)n/2$ for this model) and vary nonlinearly with
the order of the moment.

Let us write down the evolution equation for the $n$-point
correlator of a statistically stationary passive scalar field in the
frame of Kraichnan's model. The equation is
\begin{eqnarray}
&&M_n \langle\theta({\bf x}_1)\cdots\theta({\bf
x}_j)\cdots\theta({\bf x}_k)\cdots\theta({\bf x}_n)\rangle=
\frac{1}{2}\sum_{j,k=1}^n C(r_{jk}/L)\times\nonumber\\
&&\;\;\;\;\;\;\;\;\;\;\;\;\;\;\; \langle\theta({\bf
x}_1)\cdots\theta({\bf x}_{j-1})\theta({\bf
x}_{j+1})\cdots\theta({\bf x}_{k-1}) \theta({\bf
x}_{k+1})\cdots\theta({\bf x}_n)\rangle\,. \label{moment}
\end{eqnarray}
The white-in-time character of the advecting flow yields no
dependence of the $n$-th order scalar moment on mixed
velocity-scalar moments of order $n+1$, which would be a
manifestation of the well-known closure problem characteristic of
mixing in Navier-Stokes turbulence. The operator $M_n$ contains,
besides the Laplacian diffusion, an additional turbulent diffusion
part, $d_{ij}$, and can be written as
\begin{equation}
M_n=-\kappa \sum_{l=1}^n{\bf\nabla}_{{\bf x}_l}^2 +\frac{1}{2}
\sum_{l,m=1}^n d_{ij}({\bf x}_{l}-{\bf x}_m) \nabla_{{\bf x}_l}^i
\nabla_{{\bf x}_m}^j\,. \label{momentN}
\end{equation}
The so-called zero modes in the Kraichnan model are solutions of the
homogeneous subproblem in Eq.\ (\ref{moment}), i.e. when the right
hand side is set to zero. The universality of the scaling of
$S_n(r)$ with respect to $r$ is caused by the scaling dominance of
the zero modes in comparison to particular solutions of the
inhomogeneous problem (\ref{moment}). An important new insight
obtained from Kraichnan's model is indeed that zero modes are the
reason for the deviations from the classical scaling of passive
scalar structure functions in the inertial range. We wish to stress,
however, that no explicit expression for the zero modes has been
calculated, and that only expansions in limiting cases for $\zeta$
and for $\kappa\to 0$ have been found (e.g., Bernard 2000, Shraiman
\& Siggia 2000, Arad {\it et al.} 2001). However, accurate numerical
solutions have been obtained for the general case (Frisch {\it et
al.} 1998, Gat {\it et al.} 1998, Chen \& Kraichnan 1998).

The basic physical picture is this: To understand the scaling
exponent $\xi_3$ of a third-order quantity, say, it is obvious that
one needs to study the properties of objects generated from the
scalar variable at three different positions in space. At any point
in time, the three tracer particles at the three positions form a
triangle. The triangle is described by the lengthscale $R$---which,
for specificity, can be taken as the geometric mean of the lengths
of the sides of the triangle---and two of the three angles of the
triangle, say $\psi$ and $\phi$ (Pumir {\it et al.} 2000, Celani \&
Vergassola 2001). As the three particles advect, the triangles
change in shape and size. If we rescale the triangles to the same
size at each time step, the dynamics reduces to the evolution of
shapes of triangles, or to a suitable function $f(\psi, \phi)$ of
the two angles $\psi$ and $\phi$. The important result obtained for
the Kraichnan model is that the three-point statistics are governed
by those trajectories for which the change in the length scale $R$
is compensated by the change in shape of the triangles such that the
product $R^{\xi_3}f(\psi,\phi)$ is a constant. As particles move in
the Kraichnan flow, an $n$-particle cloud grows in size but
fluctuations in the cloud shape decrease in magnitude. The latter
happens because the correlation between particles---which arises
because they are contained within the integral scale of the velocity
field---weakens with the separation distance. Therefore, as
mentioned above, one looks for suitable functions of size $and$
shape that have the property of being conserved via the balance
between the growth in size and the decrease of shape fluctuations.

The important qualitative lesson from this work is that certain
types of Lagrangian characteristics, conserved only on the average,
determine the statistical scaling of Eulerian structure functions
(Falkovich \& Sreenivasan 2006). This is a clear instance where the
Lagrangian point of view has been essential to understanding better
the Eulerian quantities in turbulence---a conclusion that may have
broad validity in systems with strong fluctuations. There is
therefore enough justification for the enthusiasm about the progress
made.

However, there are several differences between the predictions of
the Kraichnan model and the behavior of a passive scalar in
Navier-Stokes turbulence. A partial list now follows:

{\bf (i)} Before we compare experiments and simulations in
Navier-Stokes turbulence with the Kraichnan model, we should
consider the meaning of the Schmidt number $Sc$ in the model. This
is not obvious because the advecting flow has no timescale for
comparison with the diffusion time.

In Navier-Stokes turbulence the smallest length is commonly thought
to be the Kolmogorov scale $\eta_K$ and the corresponding timescale
to be $\tau_{\eta} =\eta_K^2/\nu$. For $Sc\gg 1$, the passive scalar
contains scales smaller than $\eta_K$, which are acted upon by the
strain-field in the sub-Kolmogorov range, which is the same as that
imposed at $\eta_K$. Thus, the diffusive time scale
$\tau_D=\eta_B^2/\kappa$, defined by means of the Batchelor
diffusion length $\eta_B$, is the same as $\tau_\eta$, leading to
the formula $Sc=\eta_K^2/\eta_B^2$. Formally, one can define a
``Kolmogorov scale", $\tilde{\eta}_K = (\nu/D_1)^{1/\zeta}$, and a
``Batchelor scale", $\tilde{\eta}_B = (\kappa/D_1)^{1/\zeta}$, such
that
\begin{equation}
\tilde{Sc}=\left(\frac{\tilde{\eta}_K}{\tilde{\eta}_B}\right)^{\zeta}\,,
\end{equation}
can be regarded as the generalized Schmidt number (E \&
vanden-Eijnden 2001). Here, $D_1$ is  the constant in (\ref{Kraichnandiffusivity}).
This definition coincides with the classical
definition for $\zeta = 2$. We recall, however, that the Kraichnan
model discusses the advection of a scalar of fixed diffusivity in a
prescribed synthetic flow. For the latter, no knowledge of viscosity
is necessary. Therefore, despite the ingenuity of the above
argument, a proper Schmidt number does not arise as a physical
dimensionless parameter in this setting, in contrast to the
Navier-Stokes case.

{\bf (ii)} For the case of decaying turbulence behind heated grids
(for which the velocity and scalar fields are both nearly
homogeneous and isotropic, and $Sc = O(1)$), it is almost certainly
true that the decaying scalar assumes a self-similar form (i.e., the
PDF reaches a self-similar state). It is almost certainly Gaussian
(e.g., Sreenivasan {\it et al.} 1980), whereas all indications are
that they attain exponential tails for the Kraichnan model
(Balkovsky \& Fouxon 1999).

{\bf (iii)} The PDF of the scalar field in a statistically
stationary Navier-Stokes flow field with stochastic scalar driving
is almost certainly Gaussian or sub-Gaussian (Mydlarski \& Warhaft
1998, Watanabe \& Gotoh 2004). A mean scalar gradient driving yields
nearly Gaussian PDFs (Overholt \& Pope 1996, Ferchichi \& Tavoularis
2002, Schumacher \& Sreenivasan 2005) and, in some experiments, to
exponential tails (Jayesh \& Warhaft 1991, Warhaft 2000). If the
Gaussian result is correct, it would be in disagreement with the
Kraichnan model, for which the tails of the PDF are always
super-Gaussian or exponential (Shraiman \& Siggia 1994, Balkovsky \&
Fouxon 1999).

At present, we do not fully understand the circumstances under which
the passive scalar mixed by Navier-Stokes turbulence assumes a
Gaussian or exponential PDF. The shape of the PDF in inhomogeneous
shear flows is understood even less well. For example, if one
measures it on the centreline of the wake of a heated cylinder, the
PDF in the far-field has an exponential shape for the cold part
(coming from the entrainment of the ambient cold fluid) but is
Gaussian for the hot part (coming from upstream all the way from the
heated cylinder). See Kailasnath {\it et al.} (1993).

{\bf (iv)} Consider scalar gradient statistics for the Kraichnan
model. An analytical result for the tails of the PDF of scalar
dissipation $\epsilon_{\theta}=\kappa(\nabla\theta)^2$ exists for a
smooth white-in-time flow. Using the Lagrangian approach, Chertkov
{\it et al.} (1998) and Gamba \& Kolokolov (1999) deduced the
behavior to be
\begin{equation}
p(\epsilon_{\theta}){\sim}\frac{1}{\sqrt{\epsilon_{\theta}}}
\exp\left(-\epsilon_{\theta}^{1/3}\right)\;\;\;\;\;\;\mbox{for}
\;\;\;\epsilon_{\theta}\gg \langle\epsilon_{\theta}\rangle\,.
\label{pdfeps}
\end{equation}
The numerical data of Schumacher {\it et al.} (2005) from very
finely resolved simulations of high-$Sc$ mixing converge to this
formula from below.

{\bf (v)} It is still unclear in the Kraichnan model as to which
qualitative and quantitative difference arise from the finite-time
correlation of the advecting flow. This question has been addressed
at least partially in Boffetta {\it et al.} (2004). It is shown
there that  finite-time correlations of the velocity field in a
free-slip surface are important for the clustering tendency of
Lagrangian tracers, and the resemblance to the case of
infinitesimally small correlation times is only qualitative. It
needs to be shown that an instantaneously reshuffled flow can cause
the same ramp-cliff structures which are observed for advection in
Navier-Stokes turbulence. The simulations by Chen and Kraichnan
(1998) suggest that ramp-cliff features are possible even in the
Kraichnan model, even if it may appear counter-intuitive {\it a
priori}.

{\bf (vi)} The relation between active and passive scalars remains
unclear as a general principle. Celani {\it et al.} (2004) review
the work on a number of passive and active scalar fields, and
conclude that the scaling of these various fields are, in fact,
non-universal. The differences are attributed to the correlation
between the input to the scalar field and the particle trajectories,
again invoking Lagrangian interpretation.

Finally, one may speculate that the research on the Kraichnan model
has some implications for Navier-Stokes turbulence as well. It is
worth recalling that Kolmogorov formulated his original 1941 theory
in what is now called ``quasi-Lagrangian frame" (Belinicher \& L'vov
1987); it may thus be said that the importance of the Lagrangian
nature of turbulent energy cascade was thus implicit in Kolmogorov's
work. This issue was recognized also by Kraichnan (1964), who then
reformulated the DIA accordingly (Kraichnan 1965). There is still a
chasm that needs to be bridged between the work on the Kraichnan
model for the scalar and the calculation of anomalous exponents for
the hydrodynamic field (see, e.g., Chen {\it et al.} 2005), but some
progress is being made (e.g., Angheluta {\it et al.} 2006). The
nonlinearity of the Navier-Stokes equations leads to a strong
coupling of the equations of the correlations of different order,
which makes it more difficult to calculate scaling exponents from
the infinite set of equations (L'vov \& Procaccia 1998).

\section{Lagrangian simulations and experiments in Navier-Stokes turbulence}
As explained earlier, following a cloud of Lagrangian particles has
given new insights, and it is no surprise that numerical schemes
based on Lagrangian particles have been done for computing the
scaling exponents accurately (Frisch {\it et al.} 1998, Gat {\it et
al.} 1998). These calculations are not meant to be efficient for
computing all aspects of Lagrangian hydrodynamics. Indeed, for such
purposes, the preferred method is still the Direct Numerical
Simulation (DNS) of the Eulerian equations followed by smooth
interpolations of velocities at the positions of the advected
tracers (Yeung 2002, Toschi \& Bodenschatz 2009). 
The Lagrangian data are thus more expensive.
From Table I, one can infer the computational work (which is the
product of the grid points and the number of integration time steps)
needed for computing homogeneous and isotropic turbulence in a
periodic box of $N$ grid points on the side. It is typically of the
order of $N^3$; see (3.1). Here, $R_\lambda$ is the Taylor
microscale Reynolds number, $L/\eta_K$ is the ratio of the large
scale of the velocity to the Kolmogorov dissipation scale, and
$T/t_\eta$ is the ratio of the characteristic time of the large
scale $L$ to that of the dissipation scale $\eta_K$.

In Table I, except for the last row,\footnote{The last row is an
estimate of the upper bound of what is computable in principle. The
physical size of a computer cluster has been assumed to be that of
the present Earth Simulator, the size of the computing element has
been replaced by atomic dimensions (since it cannot get any
smaller!), and different parts of the computer are assumed to
communicate at the speed of light (since it cannot get any faster!).
While these upper bound estimates can be improved in detail, they
provide a reasonable order of magnitude.} the data have been deduced
from existing DNS data, but those for the Lagrangian timescale ratio
have been obtained by extrapolating existing experience at
substantially lower Reynolds numbers.

\begin{table}

\caption{Calculated and estimated values of the length and time
scale ratios in the DNS data. $N$ is the number of grid points on a
linear side of the computational box. Some of the numbers have been
taken from the simulations by Yeung (2002) and Yeung {\it et al.}
(2005).}

\vspace{0.5 truecm}
\begin{tabular}{cccc}
\hline
$N$  &$R_\lambda$&   $L/\eta$& $T/t_\eta$ \\
\hline
   128  &    89  &    56  &    8   \\
   256  &   140  &    100 &    13  \\
   512  &   230  &    190 &    20  \\
   1024 &   390  &    300 &    40  \\
   2048 &   700  &    600 &    65  \\
   4096 &   1200 &    1000&    120 \\
132,072 &   8900 &   11,000&   900  \\
\hline
\end{tabular}
\end{table}

In Eulerian turbulence, the inertial scaling range is roughly about
1/100th of $L/\eta_K$, as was discussed for example by Sreenivasan
\& Dhruva (1998) for the atmospheric boundary layer data at
$R_\lambda$ = 10,000-20,000. (Actually, this fraction appears to
depend weakly on the Reynolds number, but we shall not discuss this
detail here.) If the same factor holds for time scales as well, we
may find a decade of scaling only for $R_\lambda$ of the order of
10,000. Even this is not certain because the Lagrangian events are
distributed with stronger tails than Eulerian events (La Porta {\it
et al.} 2001, Mordant {\it et al.} 2001) and hence the incursion of
the nonuniversal features of the large scale may be stronger for
Lagrangian properties; in any case, it is clear that far higher
Reynolds numbers are needed to observe universality in Lagrangian
data. It is thus no surprise that even Richardson's law of
dispersion (Richardson 1926) is yet to be observed directly over a
decent range of scales in simulations and experiments (Ott \& Mann
2000, Boffetta \& Sokolov 2002, Biferale {\it et al.} 2005, Bougoin
{\it et al.} 2006, Schumacher 2008). Recall also that Richardson's
own compilation consisted of data from several disparate sources
and, on hindsight, leaves room for improvement. Further difficulties
in observing a Richardson scaling are related to the sensitive
dependence on the initial conditions as discussed by Bourgoin {\it
et al.} (2006) and Sawford {\it et al.} (2008).\footnote{In the
Kraichnan model, a Richardson scaling of the mean-square distance
with respect to time follows for a scaling exponent $\zeta=4/3$
instead of the classical Kolmogorov-like exponent $\zeta=2/3$.}

We believe that the estimates presented in Table I may have to be
revised unfavorably because of small-scale intermittency. According
to the well-known estimate, the computational work needed for the
DNS of turbulence varies as the third power (e.g., Orszag 1973) of
the large scale Reynolds number, $Re$:
\begin{equation}
N^3\times N_T \sim \left(\frac{L}{\eta_K}\right)^4 = Re^{3}\,.
\end{equation}
Here, $N_T$ is the number of time steps for one large-scale eddy
turnover time. It has been argued recently that intermittency
renders the smallest spatial scales smaller than the Kolmogorov
scale, $\eta_K$ (Sreenivasan 2004, Yakhot \& Sreenivasan 2005,
Schumacher {\it et al.} 2007, Schumacher 2007). Thus, in the DNS of
turbulence that computes the smallest scale, the Eulerian
computational work would increase as the fourth power of the large
scale Reynolds number (Yakhot \& Sreenivasan 2005, Schumacher {\it
et al.} 2007). Pragmatically, one may be able to work with an
intermediate value between the third and the fourth power of the
Reynolds number by sacrificing a little on the very smallest scales,
but the net effect is that one can only achieve, for a given
computational box size, lower Reynolds numbers than the above
estimates suggest.

In Lagrangian simulations, the conventional estimate for the
computational work is of the order of $Re^3\,$ln$(Re)$, the
logarithmic factor arising from interpolations of the Eulerian data.
This multiplying factor is nontrivial if $Re$ is large. All
indications are that the Lagrangian data are more intermittent in
character. If so, what is the corresponding estimate of the
computational cost, in analogy to the fourth power dependence on
Reynolds number, compared with the traditional third power in
Eulerian simulations? For a further discussion of these points,
see also Yakhot (2008b).
\begin{figure}
\begin{center}
\includegraphics[angle=0,scale=0.5,draft=false]{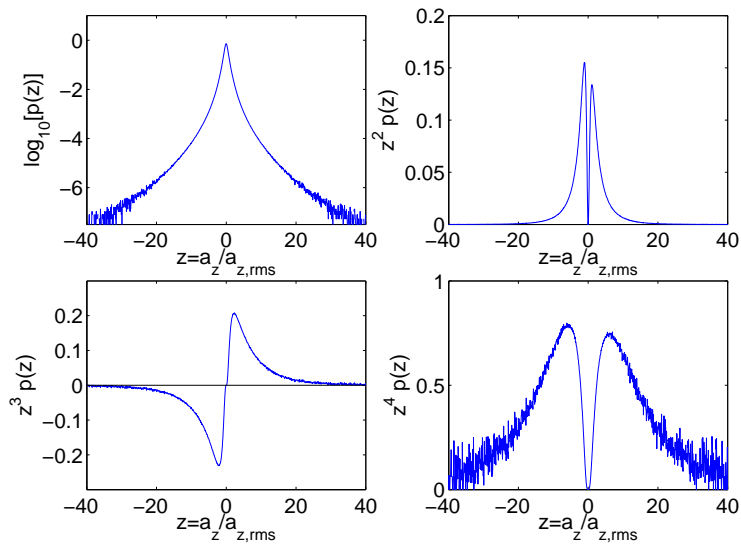}
\caption{Convergence of the statistics of the vertical acceleration
component $a_z$ in turbulent Rayleigh-B\'{e}nard convection
(Schumacher 2008, 2009). The results is gathered for $4.4\times 10^8$
events in a Cartesian slab of aspect ratio 4 which is resolved by
$2048\times 2048\times 513$ grid points. The Rayleigh number is
$Ra=1.2\times 10^8$ and the Prandtl number $Pr=0.7$. The bulk
Reynolds number is proportional to the square-root of the Rayleigh
number.} \label{fig2}
\end{center}
\end{figure}

In connection with Lagrangian intermittency (Novikov 1989), the
precise set of measurements that one should make is somewhat
unclear. In principle, the statistics of pairs of particles that
maintain a fixed separation distance will be different from those of
fixed separation along a single particle trajectory. Commonly
evaluated quantities are the moments of velocity differences of a
single Lagrangian particle taken at two times separated by a chosen
delay. The frequently used assumption of translating the
intermittency of Eulerian spatial increments to that of Lagrangian
temporal increments has been questioned (Homann {\it et al.} 2007,
Yakhot 2008b). Strong Lagrangian accelerations (i.e., strong
Lagrangian intermittency events) indeed appear frequently at the
edge of a vortex sheet or in the plane perpendicular to a
quasi-one-dimensional vortex tube, these being atypical in the
Eulerian case.

In fact, the following interesting, albeit rough, facts can be
deduced from measurements and simulations (La Porta {\it et al.}
2001, Mordant {\it et al.} 2001, Biferale {\it et al.} 2004,
Reynolds {\it et al.} 2005, Homann {\it et al.} 2007) of Lagrangian
turbulence. Lagrangian accelerations of the order of 5 times the
standard deviation occur with the frequency of one in a thousand,
those with 30 times the standard deviation occur with a frequency of
one in a million, and those of magnitude 100 times the standard
deviation occur with a non-negligible frequency of once every
billion times. It is hard to see how the non-universal effects can
be avoided altogether under the circumstances. In the experimental
data of Crawford {\it et al.} (2005), one can see that the
conditional accelerations depend quite strongly on the magnitude of
velocity fluctuations. Figure \ref{fig2} demonstrates that the
convergence of the Lagrangian statistics requires significant
efforts.  The vertical acceleration (which is less intermittent than
the lateral acceleration) in convective turbulence is shown. Even
for a record of more than $4\times 10^8$ data points,  the statistical
convergence of the fourth-order moment remains problematic.

Two consequences of these observations are worth noting. First, it
becomes more difficult to obtain converged statistics of high-order
Lagrangian moments. Second, because the tails are highly extended,
rare events contribute more strongly to high-order moments. Since
the tails are also affected by the (non-universal) large scale
features, it is not clear that one can seek universal
characteristics of high-order moments with the same level of
confidence as for Eulerian quantities. Only recently, first
successful attempts to this problem in case of Lagrangian structure
functions were reported. Theoretical predictions for the scaling
exponents in the limit $Re\to \infty$ (Zybin {\it et al.} 2008) were
found to be in remarkable agreement with experiments (Xu {\it et
al.} 2006). Certainly, more experiments are necessary in this area.
As a measure of the sensitivity of Lagrangian statistical quantities
to the tails of the distribution, one may cite the work of Sawford
{\it et al.} (2005) which shows that backward relative diffusion is
much faster than forward diffusion even for stationary and isotropic
turbulence.

\begin{figure}
\begin{center}
\includegraphics[angle=0,scale=0.45,draft=false]{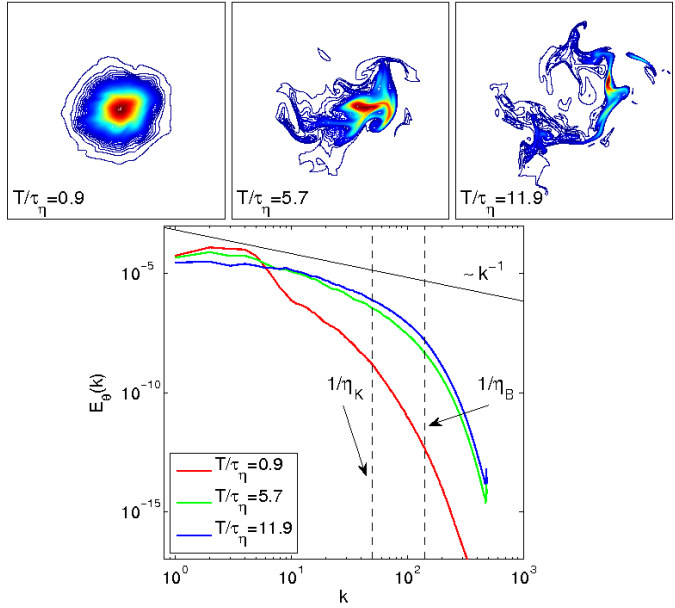}
\caption{Mixing of an initially Gaussian (and decaying) scalar blob
in a homogeneous isotropic statistically stationary turbulent
Navier-Stokes flow at a Taylor microscale Reynolds number of
$R_{\lambda}=64$ and a Schmidt number $Sc=8$. The turbulence is
resolved in a periodic box of side length $2\pi$ with $1024^3$ grid
points. The top row shows the evolution of concentration contours in
$x$-$z$ plane cuts at $y=\pi$ for progressing time. The bottom panel
shows the corresponding evolution of the scalar variance spectrum.
The classical Batchelor scaling, $E_{\theta}(k)\sim k^{-1}$, the
Kolmogorov dissipation length $\eta_K$ and the Batchelor diffusion
length, $\eta_B$ are indicated as well.} \label{fig1}
\end{center}
\end{figure}

\section{Implications for scalar mixing at high Schmidt numbers}
Let us wrap up the issues of the last two sections by discussing the
implications of the Kraichnan model for the particularly important
case of scalar mixing at high Schmidt numbers. We will consider
three aspects.

The argument can be made that the Kraichnan model for the smooth
limit of the velocity field, $\zeta\to 2$, comes close to the
high-$Sc$ mixing regime in a Navier-Stokes flow with very large
Reynolds numbers. The latter would yield a very short Kolmogorov
timescale $\tau_{\eta}$ and an extended viscous-convective range.
However, one fundamental difference remains. The finite time over
which a local flow pattern with compressional strain persists is a
necessity to steepen the frequently observed sheet-like scalar
gradient patterns against the action of molecular diffusion
(Sreenivasan \& Prasad 1989, Dahm {\it et al.} 1991, Buch \& Dahm
1996, Villermaux \& Meunier 2003, Schumacher {\it et al.} 2005).
Therefore, the permanent re-shuffling of the local flow patterns in
the Kraichnan case will destroy the well-known stretch-twist-fold
scenario, and sheet-like structure of the scalar gradient fields may
well be absent. However, statistically, the Kraichnan model does
produce very strong gradients (as given by (\ref{pdfeps}) for the
tails), which seem to form a limit to the mixing in Navier-Stokes
turbulence. All current numerical studies on high-Schmidt-number
mixing in Navier-Stokes turbulence suffer from the small Reynolds
numbers that can be obtained when the largest gradient fluctuations
are resolved (Schumacher {\it et al.} 2005). Therefore, the
existence of the so-called mixing transition, which postulates a
weaker Reynolds number dependence of scalar mixing at large Reynolds
numbers (Dimotakis 2005, Yakhot 2008a), is still an open issue.
Progress on the structural differences and their relation to the
statistics for both mixing schemes will be quite useful.

Although the closure problem is removed in the Kraichnan model (see
(2.4)), analytical predictions on the anomalous scaling of scalar
structure functions, $S_n(r)$, are possible only for limiting cases.
For example, the Kraichnan model gives a scaling exponent
$\xi_3=3-7\zeta/5$ when the scalar fluctuations are sustained by a
constant mean scalar gradient (Pumir 1996). This scaling could be
checked, e.g., by advecting the scalar in Navier-Stokes turbulence
first and determining $\xi_3$. The time correlations of the
advecting flow could be destroyed as in Boffetta {\it et al.} (2004)
in order to obtain a Kraichnan flow and the third-order moment
analysis can be repeated subsequently. Here, a direct link to the
Lagrangian viewpoint is also possible by solving the (stochastic)
equation (\ref{sde1}) for triplets or quadruplets of tracers and
studying shapes in the spirit of Pumir {\it et al}. (2000) and
Celani \& Vergassola (2001).

Kraichnan's (1968) original motivation for his model was to
demonstrate that the opposite extreme to Batchelor's (1959)
quasistatic straining motion can result in the same scalar variance
spectrum $E_{\theta}(k)\sim k^{-1}$. This can be seen to be so from
(\ref{momentN}) which becomes, for the second order and $\zeta=2$,
\begin{equation}
-[2\kappa{\bf\nabla}_{\bf r}^2 +d_{ij}({\bf r}) \nabla_{\bf r}^i
\nabla_{\bf r}^j]\langle\theta({\bf x})\theta({\bf x}+{\bf
r})\rangle=C(r/L) \,. \label{moment2}
\end{equation}
The second term on the left hand side corresponds to the scalar
variance transfer term, $T_{\theta}(k)$, for homogeneous isotropic
turbulence and reads
\begin{equation}
T_{\theta}(k)=2 D_1 \frac{\partial}{\partial k}\left[k^4 \frac{\partial}{\partial k}\left(
\frac{E_{\theta}(k)}{k^2}\right)\right]\,,
\label{moment2k}
\end{equation}
after Fourier transformation. Indeed, $E_{\theta}(k)\sim k^{-1}$
yields a $k$-independent transfer rate $\int^{\infty}_k
T_{\theta}(p)\,\mbox{d}p$ in the viscous-convective range. The
experiments which demonstrate such spectral roll-off are quite rare;
see, e.g., Villermaux {\it et al.} (2001), who note that the
$k^{-1}$ power requires both a large range of scalar scales and a
sufficiently large Reynolds number.\footnote{A corresponding
logarithmic scaling of the second order structure function was,
however, observed by Borgas {\it et al.} (2004).} It is furthermore
unclear how the large-scale forcing of the passive scalar will
affect the spectrum. Figure \ref{fig1} illustrates a simulation
which seems to be the simplest and therefore most transparent case
for verifying the concepts of both Batchelor (1959) and Kraichnan
(1968). A scalar concentration blob is advected in a statistically
stationary turbulent flow for $Sc=8$. The data correspond only to
moderate Schmidt numbers because of the current numerical
limitations, but perhaps display a slow approach to the $k^{-1}$
spectrum.

\section{Concluding remarks}

The recent Lagrangian work has ushered in a breath of fresh air in
turbulence, but it has not been sufficiently integrated with prior
Lagrangian perspectives. Among others, most of the recent work has
focused exclusively on material particles, but there are many other
Lagrangian aspects to the turbulence problem. For example, classical
works of Taylor \& Green (1937) and Taylor (1938) discuss vortex
line-stretching as the basic Lagrangian mechanism of turbulence. The
modern work has little to offer towards understanding that problem.
The general problems raised by Batchelor (1952)---such as the
extension of material lines and surfaces and fluxes across
surfaces---have yet to be understood well, even for the Kraichnan
model; even basic issues such as the existence of material lines and
surfaces in the infinite Reynolds number limit have yet to be
addressed. The list of unanswered questions might also include the
fractal nature of isosurfaces (e.g., Sreenivasan 1991). It would be
useful to answer questions such as: what are fractal dimensions of
isoscalar surfaces and other material objects in the Kraichnan
model? There is extensive numerical work on the problem (see San Gil
2001), but no theoretical answers. This particular question is
important for the mixing of reactive scalars or turbulence in
clouds. Scalar isolevel sets in Navier-Stokes case do not display
monofractal behaviour for low-Reynolds-number flows (Frederiksen
{\it et al.} 1997, Schumacher \& Sreenivasan 2005), though it
appears that larger Reynolds numbers of the advecting flow will
change this behavior. Other approaches such as the dissipation
element analysis which reduces the mixing to a permanent reshuffling
of the separatrix lines between zeros of the scalar gradient might
also provide complementary insight on the nature of high-Sc mixing
(Wang \& Peters 2006).

It is our belief that the Lagrangian perspective on turbulence is
really essential and that it has not been pushed far enough. It is
also our belief that the intrinsic mechanisms of turbulence are
essentially Lagrangian.

{\em Acknowledgements.} We wish to thank Werner Dahm, Gregory Eyink,
Brian Sawford, Emmanuel Villermaux, Victor Yakhot and
P.K. Yeung for helpful discussions. This work was supported by the German Academic
Exchange Service (DAAD), the Deutsche Forschungsgemeinschaft (DFG)
within the Heisenberg Program and the US National Science Foundation
(NSF). Computing resources and support from the J\"ulich
Supercomputing Centre (Germany) under grants HMR09 and HIL02 are
gratefully acknowledged.

\end{document}